\documentclass[sn-nature]{sn-jnl}
\usepackage{graphicx}
\usepackage{multirow}
\usepackage{amsmath,amssymb,amsfonts}
\usepackage{amsthm}
\usepackage{mathrsfs}
\usepackage[title]{appendix}
\usepackage{xcolor}
\usepackage{textcomp}
\usepackage{manyfoot}
\usepackage{booktabs}
\usepackage{algorithm}
\usepackage{algorithmicx}
\usepackage{algpseudocode}
\usepackage{listings}
\usepackage{comment}
\usepackage{booktabs}
\usepackage{lineno}
\usepackage{geometry}

\theoremstyle{thmstyleone}

\theoremstyle{thmstyletwo}

\theoremstyle{thmstylethree}

\begin{document}

\title[AI decisions in human interaction]{\vspace{-2cm}\textbf{Generative AI Triggers Welfare-Reducing Decisions in Humans}}

\author*[1,2]{\fnm{Fabian} \sur{Dvorak}}\email{\small fabian.dvorak@uni-konstanz.de}
\equalcont{\small These authors contributed equally to this work.}

\author*[3,4]{\fnm{Regina} \sur{Stumpf}}\email{\small regina.stumpf@uni-konstanz.de}
\equalcont{\small These authors contributed equally to this work.}

\author[5]{\fnm{Sebastian} \sur{Fehrler}}\email{\small sebastian.fehrler@uni-bremen.de}

\author[3,4,6]{\fnm{Urs} \sur{Fischbacher}}\email{\small urs.fischbacher@uni-konstanz.de}

\affil[1]{\orgdiv{\footnotesize Centre for the Advanced Study of Collective Behaviour}, \orgname{University of Konstanz}, \orgaddress{\street{Universit\"{a}tsstrasse 10}, \city{Konstanz}, \postcode{78464}, \country{Germany}}}

\affil[2]{\orgdiv{\footnotesize Department of Environmental Social Sciences}, \orgname{Eawag}, \orgaddress{\street{\"{U}berlandstrasse 133}, \city{D\"{u}bendorf}, \postcode{8600}, \country{Switzerland}}}

\affil[3]{\orgdiv{\footnotesize Department of Economics}, \orgname{University of Konstanz}, \orgaddress{\street{Universit\"{a}tsstrasse 10}, \city{Konstanz}, \postcode{78464}, \country{Germany}}}

\affil[4]{\orgdiv{\footnotesize TWI}, \orgname{Thurgau Institute of Economics}, \orgaddress{\street{Hafenstrasse 6}, \city{Kreuzlingen}, \postcode{8280}, \country{Switzerland}}}

\affil[5]{\orgdiv{\footnotesize SOCIUM}, \orgname{University of Bremen}, \orgaddress{\street{Mary-Somerville-Straße 7}, \city{Bremen}, \postcode{28359}, \country{Germany}}}

\affil[6]{\orgdiv{\footnotesize CESifo}, \orgaddress{\street{Poschingerstraße 5}, \city{Munich}, \postcode{81679}, \country{Germany}}}

\abstract{\vspace{-0.1cm}
Generative artificial intelligence (AI) is poised to reshape the way individuals communicate and interact.
While this form of AI has the potential to efficiently make numerous human decisions, there is limited understanding of how individuals respond to its use in social interaction.
In particular, it remains unclear how individuals engage with algorithms when the interaction entails consequences for other people.
Here, we report the results of a large-scale pre-registered online experiment (N = 3,552) indicating diminished fairness, trust, trustworthiness, cooperation, and coordination by human players in economic two-player games, when the decision of the interaction partner is taken over by ChatGPT.
On the contrary, we observe no adverse welfare effects when individuals are uncertain about whether they are interacting with a human or generative AI. Therefore, the promotion of AI transparency, often suggested as a solution to mitigate the negative impacts of generative AI on society, shows a detrimental effect on welfare in our study. Concurrently, participants frequently delegate decisions to ChatGPT, particularly when the AI's involvement is undisclosed, and individuals struggle to discern between AI and human decisions.
}



\maketitle

\section{Introduction}\label{sec1}
Fairness, reciprocity, trust, cooperation, and coordination are fundamental to the proper functioning of human societies. These facets of human behavior support socially desirable outcomes by promoting efficiency through coordination, and stabilizing norms of trust, fairness, and cooperation. Recent advances in the development of generative artificial intelligence have shown the potential to fundamentally change nearly every aspect of life in modern societies. While it is clear that AI has an enormous potential for improving human life~\citep{Vinuesa2020,Lin2023,Peng2023,Wang2023,Yan2023,Lam2023,Bi2023} and increasing economic productivity~\citep{Abramoff2023,Noy2023}, little is known about the consequences of using generative AI in everyday human interaction. In this study, we approach the question of how generative AI will affect human interactions from a welfare perspective. In particular, we ask how the now ubiquitous possibility of using generative AI will affect our everyday social interactions.

Human social behavior depends not only on the expected consequences of our actions for others~\citep{Tricomi2010,Fehr2003}, but also on our beliefs about others' behavior~\citep{Fehr2004}, which jointly explain the variability of culture and social norms~\citep{Henrich2001}. If decisions in social interaction are (potentially) delegated to AI, individual's beliefs and preconceptions about the nature of AI, and about the circumstances under which AI systems are employed, will become pertinent factors influencing their decisions~\citep{Pataranutaporn2023}.

The focus of this study is \textit{not} how generative AI will behave in social interactions, which has been studied in a recent string of other papers \citep{bauer_decoding_2023,chen_emergence_2023,guo_gpt_2023,dargnies_aversion_2022}. Instead, we investigate \textit{human reactions} to the utilization of generative AI in social interactions, an aspect that has been underexplored thus far.\footnote{Research on human-algorithm interaction generally shows that, depending on the design of the algorithm, humans often act more rationally, selfish, and seem less prone to emotional and social responses~\citep{Chugunova2022,Kobis2023}, but are nevertheless willing to delegate decisions to algorithms~\citep{Candrian2022}. The key feature of our study is that the algorithm acts on behalf of a real participant, who is affected by the consequences of the interaction, which matters for social preferences~\citep{Schenk2024}. More loosely related to our study is the emergent literature investigating humans' willingness to follow algorithmic advice~\citep{kawaguchi2021will,logg2019algorithm,Greiner2022,zhou2021exploitation}.} In particular, several fundamental questions regarding human behavior in AI-mediated social interaction remain unaddressed. What are the welfare consequences of human reactions to the use of generative AI in social interaction? Under what circumstances are people willing to delegate decisions to generative AI? What role does the transparency of AI decisions play? And does the potential to personalize generative AI models influence how people react to their use?

We conduct a pre-registered online experiment (N=3,552) to investigate the repercussions of incorporating generative AI into human interaction. In the main experiment, 2,905 participants (mean age: 39.9 years, 49.6\% women, 47\% college or university degree) engage in direct interactions either with other participants or with the large-language model ChatGPT~\citep{ChatGPT} acting on behalf of a human participant with AI support.
In each interaction, one of the two participants is supported by ChatGPT. Both participants are affected by the consequences of the interaction, even if the AI is acting on behalf of the participant with AI support. We provide the same instructions to human participants and the AI and ask for their decisions together with short statements of justification in 5 standard two-player games with direct welfare consequences: the ultimatum game (UG)~\citep{Gueth1982}, the binary trust game (bTG)~\citep{Berg1995}, the prisoner’s dilemma game (PD)~\citep{Lave1962}, the stag hunt game (SH), and the coordination game (C). For all five games, a large body of empirical research exists, documenting how humans interact under controlled laboratory conditions~\citep{Johnson2011,Oosterbeek2004,Mengel2017,DalBo2021}. To incentivize the decisions, each pair of participants receives the payoffs that result from one randomly selected interaction. 

The experiment employs a between-subject design featuring six treatments. In the first treatment condition (\textit{transparent random}), the AI randomly decides on behalf of the participant with AI support with a 50\% probability. Participants without AI support make two decisions -- one for interacting with a human and another for interacting with AI -- rendering the use of AI transparent. In the second treatment condition, the participant with AI support has the option to decide whether to delegate her decision to AI (\textit{transparent delegation}). Meanwhile, the participant without AI support makes two decisions, accounting for each possible outcome of the delegation decision. The third treatment condition is a variation of the delegation treatment, where the participant without AI support cannot condition her decision on the outcome of the delegation decision, rendering the use of AI opaque (\textit{opaque delegation}). For each of these three treatment conditions, we vary whether participants can personalize the decisions AI makes on their behalf at the beginning of the experiment (\textit{personalized}) or not (\textit{non-personalized}, which results in six between-subject treatments overall. 

To personalize the decision of AI, participants answered seven binary questions about their own personality before receiving the instructions of the games. The questions followed a simple "A or B?" format with the following pairs of alternatives: Intuition or Thoughtfulness, Introversion or Extraversion, Fairness or Efficiency, Chaos or Boredom, Selfishness or Altruism, Novelty or Reliability, and Truth or Harmony. Participants knew that their answers to the 7 binary preference questions would be used to personalize the AI decisions made on their behalf. They also knew that whenever they interacted with the AI during the experiment, the algorithm would make decisions according to the other participant's preferences, which were elicited through the same questions. We sampled personalized AI decisions for each possible combination of 7 binary preferences ($2^7 = 128$) by prompting ChatGPT to generate decisions made by a person whose preferences were reflected in a particular response pattern (see Methods for details).  

In addition to the main experiment, we test if AI decisions can be detected in social interaction by showing sets of AI decisions and human decisions to 647 human raters (mean age: 39.8 years, 50.1\% women, 43.0\% college or university degree). Raters receive a bonus payment if they accurately identify the decisions made by the AI. We also perform Turing tests by providing human raters with written justifications for the decisions generated by humans and AI.  

\section{Experimental Results \& Discussion}\label{sec2}

To quantify the consequences of using AI in social interactions, we use the pre-registered indices shown in Table \ref{tab:indices}~\citep{OSF}. To construct the indices, we sum up signed normalized decisions in the games, and use the average of the individual averages. In the welfare index, all decisions enter positively.\footnote{The only decision for which a positive relation with welfare is debatable is the minimum acceptance threshold in the ultimatum game. Using a positive weight requires that long-term gains of upholding the social norm of utility will outweigh the short-term efficiency losses of rejections. Our results get even stronger if we use a negative weight for the responder decision.} Three additional indices quantify the impact of prosociality, reciprocity, and beliefs about the kindness of the other player, revealing their influence on the underlying welfare effects.

The main hypothesis tests that we present in Table \ref{tab:main_results} were preregistered.\footnote{The reported p-values are adjusted for multiple testing using the Holm-Bonferroni method \citep{Holm1979}. This explains why some of the reported p-values are exactly equal to 1. The alpha level to define statistical significance is 5\%.} The main results are consolidated within the same table, delineating the research questions and their corresponding answers derived from our experimental results for each inquiry. The table specifies the tested variable and the compared treatment conditions.

\begin{table}[h]
\caption{Behavioral indices}\label{tab:indices}
\begin{tabular}{|l|p{.30\textwidth}|p{.45\textwidth}|}
\toprule
Index & Description & Game Decisions (effect) \\
\midrule
Welfare index &  Combines all decisions with welfare consequences. & normalized offer in UG (+), normalized minimum acceptance threshold in UG (+), binary trust decision in bTG (+), normalized back-transfer in bTG (+), cooperation in PD (+), cooperation in SH (+), modal choice in C (+)\\
\midrule
Prosociality index & Combines decisions in which social preferences play a role. Prosocial decisions are defined as decisions that increase the (expected) payoff of the other participant. & normalized offer in UG (+), normalized minimum acceptance threshold in UG (--), normalized binary trust decision in bTG (+), normalized back-transfer in bTG (+), cooperation in PD (+), cooperation in SH (+)\\
\midrule
Kindness index & Quantifies the beliefs in the kindness of the other player. & normalized binary trust decision in bTG (+), cooperation in PD (+), cooperation in SH (+).\\
\midrule
Intentions index & Quantifies the role of intentions. & normalized minimum acceptance threshold in UG (+), normalized back-transfer in bTG (+). \\
\bottomrule
\end{tabular}
\footnotetext{Note: Behavioral indices created from the normalized game decisions. All indices were pre-registered (https://osf.io/fvk2c/).}
\end{table}

The key result of our study is that engaging with generative AI triggers human decisions that lead to a decrease in welfare (Question 1). This result emerges consistently across all five experimental games, with significant consequences for participants' payoffs. We further find that participants often delegate their decisions to generative AI, especially, if delegation is opaque (Question 2). 
Surprisingly, we observe that participants do not alter their behavior when delegation is opaque (Question 3), despite their anticipation of others delegating. Delegation does not crowd-out the social preferences of individuals who are aware that the other person delegated to AI (Question 4). We find that personalizing the AI model does not affect how people respond to it (Questions 5-8). In particular, it does not restore the welfare loss caused by the interaction with AI. We will now delve into each question in more detail. Initially, we present the results from the non-personalized treatments and subsequently highlight the impact of personalization.
\setlength{\tabcolsep}{1pt}
\begin{table}[h]
\caption{Key findings of the main experiment}\label{tab2}
\label{tab:main_results}
\begin{footnotesize}
\begin{tabular*}{\textwidth}{@{\extracolsep\fill}lllrrr}
\toprule%
Research question & & Variable & Cond. 1 & Cond. 2 & p-val  \\
\midrule
1. Does interacting with AI decrease welfare? &\textit{Yes}& welfare index & HU: 0.69 & AI: 0.65 & $< 0.001$ \\
2. Is opaque delegation more frequent? &\textit{Yes}& delegation freq. & TD: 0.38 & OD: 0.42 & $0.006$\\
3. Does opaque interaction decrease welfare? &\textit{No}& welfare index & HU: 0.69 & UN: 0.71 & $1.000$ \\
4. Does delegation crowd-out prosociality? &\textit{No}& prosociality index & R: 0.55 & D: 0.53 & $0.422$\\
\midrule
\multicolumn{5}{@{}l@{}}{Does personalizing the AI...} \\
\midrule
5. ...restore welfare? & \textit{No} & welfare index & P: 0.66 & NP: 0.65 & $1.000$\\
6. ...increase delegation? & \textit{No} & delegation freq. & P: 0.39 & NP: 0.40 & $1.000$\\
7. ...change welfare if delegation is opaque? & \textit{No} & welfare index & P: 0.70 & NP: 0.71 & $1.000$\\
8. ...change prosocial behavior? & \textit{No} & prosociality index & P: 0.55 & NP: 0.53 & $0.422$\\
\botrule
\end{tabular*}
\footnotetext{Note: All hypothesis tests and indices were pre-registered and controlled for multiple testing (https://osf.io/fvk2c/). Results of one-sided t-tests with Holm-Bonferroni corrected p-values~\citep{Holm1979}. Corrected p-values can be equal to 1. HU: decisions directed at human participant; AI: decisions directed at AI; UN: unknown interaction partner (AI or human); TD: transparent delegation; OD; opaque delegation; R: random takeover by AI; D: delegation; P: personalized AI; NP: non-personalized AI.}
\end{footnotesize}
\end{table}

\paragraph{Does the use of generative AI in human interaction decrease welfare?}
When interacting with the AI, participants' decisions lead to a significantly lower level of welfare (M = 0.65, SD = 0.19) than when interacting directly with a human (M = 0.69, SD = 0.17, t(482) = -5.48, p $<$ 0.001). This is also true if the other person has intentionally delegated the decision to the AI (AI: M = 0.63, SD = 0.20, HU: M = 0.71, SD = 0.17, t(480) = -10.19, p $<$ 0.001) and if the AI is personalized to reflect the preferences of the other person (AI: M = 0.63, SD = 0.20, Human: M = 0.71, SD = 0.17, t(1902) = -15.42, p $<$ 0.001). 
Most of the components of the welfare index contribute to this result.\footnote{We pre-registered to analyze the components of the significant main indices.} Beliefs in the kindness of the other player are significantly reduced when interacting with AI (M = 0.61, SD = 0.32) compared to interacting with humans directly (M = 0.66, SD = 0.31, t(482) = -3.78 , p $<$ 0.001). 
The frequency of the modal choice in the coordination game (earth), which we use as a measure for the predictability of the participants' decision, is significantly lower when interacting with AI (AI: M = 0.57, SD = 0.50 , HU: M = 0.64, SD = 0.48 , t(482) = -2.73 , p = 0.003). 
The normalized offer in the ultimatum game, which can be seen as an incentivized measure for equality concerns is also significantly lower when interacting with AI (M = 0.91, SD = 0.31) compared to direct interaction with a human (M = 0.98, SD = 0.24, t(482) = -5.02, p $<$ 0.001). Reciprocity does no seem to play a role in explaining the welfare losses when interacting with AI (intention index, AI: M = 0.62, SD = 0.19, HU: M = 0.63, SD = 0.19, t(482) = -1.27, p = 0.102).

The reduction in welfare is also reflected in lower levels of expected payoffs of participants when interacting with AI (see bottom panel in Figure \ref{fig:game_results_with_payoff}). The simulated expected payoff of all matched decisions that were targeted at the AI (M = 4.74, SD = 0.36) is significantly lower than the one where decisions were targeted at humans (M = 5.06, SD = 0.34, t(1902) = 43.79, p $<$ 0.001). Thus, the involvement of AI in human social interactions leads to an overall loss in expected average payoff of 6\%. This discrepancy in expected average payoff varies from a 3\% loss in the prisoner's dilemma to a 10\% loss in the ultimatum game. 

We use linear regression models to explore the extend to which the AI-triggered welfare loss is explained by participants' prior experience with ChatGPT, their socio-economic characteristics, and their attitudes towards AI (models (4)--(6) of Table~\ref{tab:regression}). We find that prior experience with ChatGPT does not significantly reduce the welfare loss. Age and positive attitudes towards AI, like the belief that AI is trustworthy, and the equality concern in an interaction with AI, mitigate the welfare loss, while the subjective difficulty of predicting AI decisions increases it. Participants that took longer in the experiment show a higher welfare loss.\footnote{The size of the AI-induced welfare effect is also robust when controlling for socio-economic characteristics of the participants, their prior usage of ChatGPT, and their attitudes towards AI (see models (1)--(3) in Table~\ref{tab:regression}).} 

\paragraph{Is opaque delegation more frequent?}
We find that the propensity to delegate the decision to AI increases when delegation is opaque. Comparing the delegation frequency in the treatments with transparent delegation (M = 0.38, SD = 0.29) to the delegation frequency in the treatments with opaque delegation (M = 0.42, SD = 0.29) shows that the propensity to delegate increases by 4 percentage points (t(1944) = 3.14, p = 0.006). This might be explained by the fact that participants anticipate the negative welfare consequences when the use of AI is transparent, and shy away from transparent delegation. Participants' answers to a post-experimental question indicate that transparent delegation is generally considered less appropriate than opaque delegation (transparent delegation (0-4): M = 2.47, SD = 0.97, opaque delegation (0-4): M = 2.61, SD = 0.92, t(1920) = 3.08, p = 0.002). 

\paragraph{Does opaque delegation decrease welfare?}
Unexpectedly, we find that welfare is not affected if opaque delegation to AI is possible. The decisions of participants who do not know if they interact with a human or AI produce a similar level of welfare (M = 0.71, SD = 0.17) compared to the decisions of participants who know that they interact with a human (M = 0.69, SD = 0.17, t(968) = 1.46, p = 1.000). At the same time, participants frequently delegate their decisions to AI (M = 0.42, SD = 0.29) if delegation is opaque and also expect others to use opaque delegation according to a post-experimental question (belief in delegation (0-4): M = 2.85, SD = 0.86). 

\paragraph{Does delegation to AI crowd-out prosocial behavior?} 
Comparing the participants' decisions when interacting with AI in situations in which the decision was delegated to AI with situations in which AI took over randomly allows us to study participants' reactions to delegation. We find that delegation does not crowd out social preferences. Prosocial behavior is unchanged in situations when delegation took place (prosociality index: M = 0.53, SD = 0.19) compared to situations when the AI took over the decision randomly (M = 0.55, SD = 0.20, t(960) = -1.48, p = 0.422). The finding is further substantiated by participants' answers to a question in the post-experimental questionnaire, which indicates that transparent delegation is generally considered appropriate (appropriateness (0-4): M = 2.47, SD = 0.97, comparison to indifference (2), t(945) = 14.98, p $<$ 0.001).

\paragraph{Does personalization of the AI matter?}
We find that personalization of AI does not restore the welfare losses that result from the usage of AI in social interaction. The welfare consequences of participants' decisions are similar in the non-personalized random treatment (M = 0.65, SD = 0.19) and the personalized random treatment (M = 0.66, SD = 0.19, t(952) = 0.57, p = 1.000). This also applies in the treatments with opaque delegation (non-personalized: M = 0.71, SD = 0.17, personalized: M = 0.70, SD = 0.18, t(997) = -1.15, p = 1.000). Neither does personalization increase the propensity to delegate to AI (non-personalized: M = 0.40, SD = 0.29, personalized: M = 0.39, SD = 0.29, t(1946) = -0.85, p = 1.000). Responses to our post-experimental question reveal that the utilization of personalization has a noteworthy impact on enhancing the alignment between the AI's decisions and those of the other individual, as indicated by participants' responses (non-personalized (0-4): M = 1.54, SD = 1.10 , personalized (0-4): M = 1.94, SD = 1.12, t(2902) = -9.74, p $<$ 0.001). However, participants also indicate that the other person is not adequately represented by the personalized AI (AI adequately represents human (0-4): M = 1.94, SD = 1.12, comparison to indifference (2), t(1451) = -2.07, p = 0.981). \\

\subsection{Experimental Games}
\begin{figure}[htbp]%
\centering
\includegraphics[width=\textwidth]{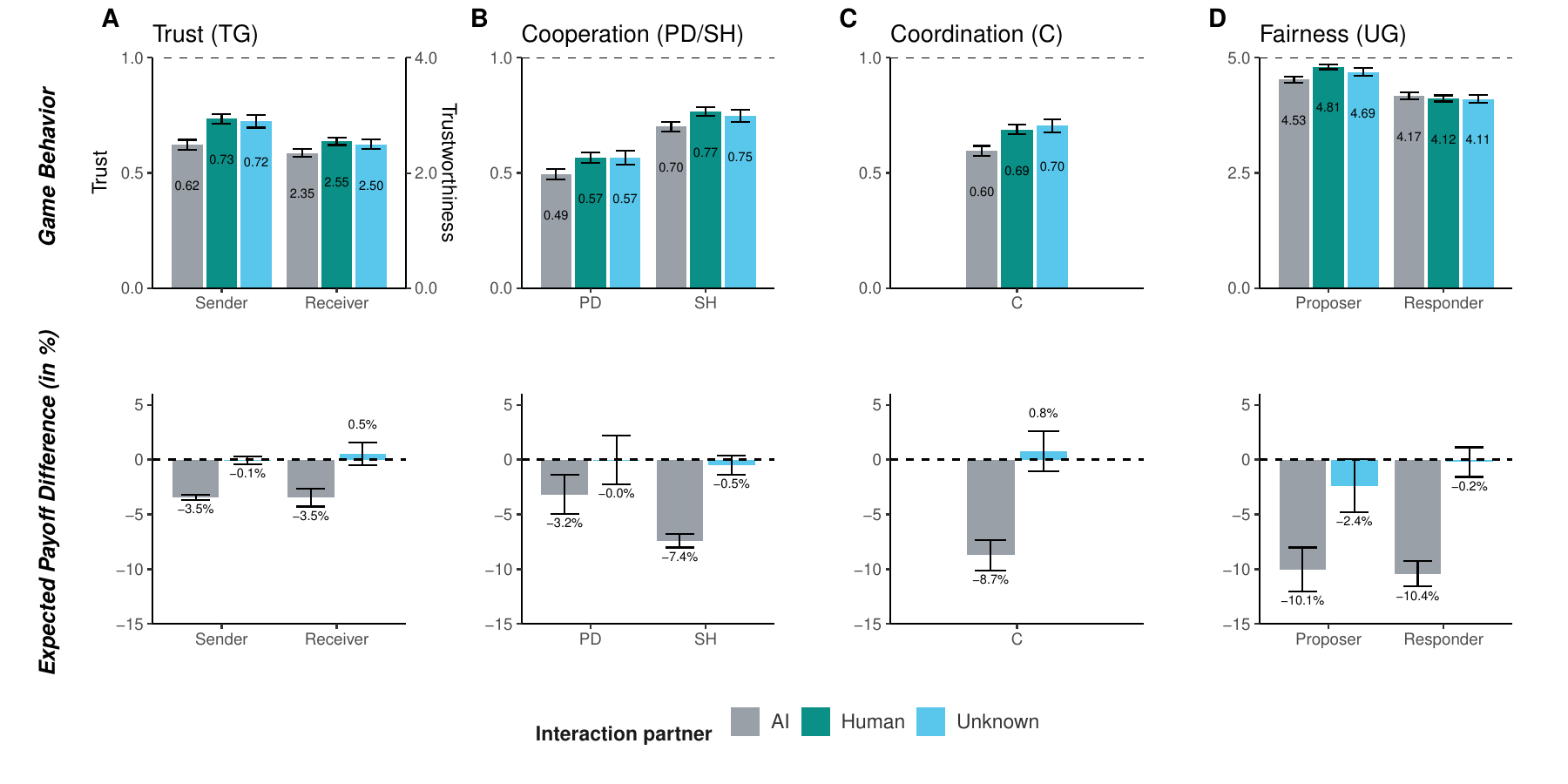}
\caption{Results across four types of games: Trust Game (Panel A), Cooperation Games (Panel B), Coordination Game (Panel C), Fairness Game (Panel D). The color of the bars represent decisions in which the interaction partner is AI (gray bars), human (green bars) or unknown (blue bars). \\
Top row: average decision in each game. Bottom row: simulated expected payoff difference. Each participant’s decisions targeted at humans, AI
or unknown is matched to the respective decisions of all other participants and the mean payoff is computed for each participant and each situation resulting from all matched interactions. Expected relative payoff difference
to human interactions are reported.
Whiskers show bootstrapped 95\% confidence intervals.}\label{fig:game_results_with_payoff}
\end{figure}

The graphs in the top row of Figure~\ref{fig:game_results_with_payoff} show the average decision in each game for three cases: the case where the interaction partner is human (green bars), AI (gray bars) or unknown (blue bars). The bottom graph shows the relative expected payoff differences for two cases where the interaction partner is AI or unknown compared to the case where the interaction partner is human.\footnote{We compare the average payoff for human interactions (where both humans knowingly play against a human) to the average payoff for human interactions in which both humans believe to interact with AI or an unknown interaction partner respectively. We match each participant's decisions targeted at humans, AI or unknown to the respective decisions of all other participants and compute for each participant the mean payoff for each situation resulting from all matched interactions. We report the expected mean difference in payoffs comparing it to the baseline payoff resulting when the interaction partner is human.}
Whiskers show 95\% confidence intervals resulting from non-parametric bootstrapping.

Comparing the decisions directed to the AI with the decisions directed to other humans, we see welfare-decreasing decisions in all five experimental games. 
Panel A shows that participants place less trust in the AI (M = 0.62, SD = 0.48) than in humans (M = 0.73, SD = 0.44), and are less trustworthy if the trust decision was made by AI (M = 2.35, SD = 1.50) than when it was made by a human (M = 2.55, SD = 1.38). Both effects together imply a payoff loss of 3.5 percent in each role of the trust game.

Panel B shows that participants cooperate less frequently in the prisoner's dilemma when the other player's decision is taken by the AI (M = 0.49, SD = 0.50) than when taken by a human (M = 0.57, SD = 0.50), which decreases the average payoff in this game by 3.2 percent. The same result applies to the stag-hunt game (AI: M = 0.70, SD = 0.46; HU: M = 0.77, SD = 0.42), leading to a payoff loss of 7.4 percent in this game. 
Panel C shows that the predictability of participants' actions in the coordination game suffers when interacting with the AI (M = 0.60, SD = 0.49) compared to the human interaction (M = 0.69, SD = 0.46), resulting in a payoff loss of 8.7 percent. 

Panel D shows that participants offer less as proposers in the ultimatum game if they know that the decision of the responder is taken over by the AI (M = 4.53, SD = 1.56) compared to when the responder herself takes the decision (M = 4.81, SD = 1.22). When the AI makes the offer in the ultimatum game, participants increase their minimum acceptance threshold and tolerate less negative inequality (M = 4.17, SD = 1.62) compared to when the offer is made by a human (M = 4.12, SD = 1.47). This is the only situation in which the reaction to the AI is potentially welfare increasing in the long run, as participants are willing to uphold the fairness norm at personal costs.\footnote{Our welfare index relies on the notion that the benefits of upholding a social norm of equality will outweigh the efficiency loss in the long run. A higher minimum acceptance threshold is therefore welfare increasing according to our welfare index. It should be noted that modifying our welfare index by considering higher minimum acceptance thresholds as welfare decreasing will make the welfare losses induced by generative AI even stronger. Welfare is also significantly reduced when excluding the ultimatum game from the welfare index.} At the same time, the larger minimum acceptance threshold has negative consequences for efficiency as it increases the likelihood of rejections. Combined with the lower offers, the interaction with AI reduces the expected payoffs of the proposer and the responder by 10.1 percent and 10.4 percent respectively.

The welfare-reducing choices made when interacting with AI are substantiated by participants' answers in the post-experimental questionnaire. Participants think AI is less trustworthy and less cooperative than humans (trustworthiness and cooperation (0-4): AI: M = 2.08, SD = 0.81; Human: M = 2.25, SD = 0.71; testing difference in trustworthiness and cooperation (0): M = -0.17, SD = 0.99, t(2904) = -9.44, p $<$ 0.001), which explains reduced trust as sender in the trust game and less cooperation in the prisoner's dilemma and the stag-hunt game when the decision of the other participant is made by AI.
Participants also indicate that they care less about equality when AI makes the decision on behalf of the other person (equality concern (0-4): AI: M = 2.98, SD = 1.05; Human: M = 3.26, SD = 0.95, testing difference in equality concern (0): M = -0.28, SD = 0.87, t(2904) = -17.06, $<$ 0.001), which explains lower offers as proposers in the ultimatum game and lower back-transfers as receivers in the binary trust game. 

The blue bars in Figure~\ref{fig:game_results_with_payoff} show that participants decisions in situations in which they do not know if they interact with a human or the AI are similar to their decisions directed at humans (green bars). This is the case for every decision in each game, and suggests that, if in doubt, participants behave as if they would directly interact with a human. Consequently, there is no expected payoff difference between human interactions and interactions in which the interaction partner (AI or human) is not known.

\subsection{Delegation Behavior}
\begin{figure}[htbp]%
\centering
\includegraphics[width=0.8\textwidth]{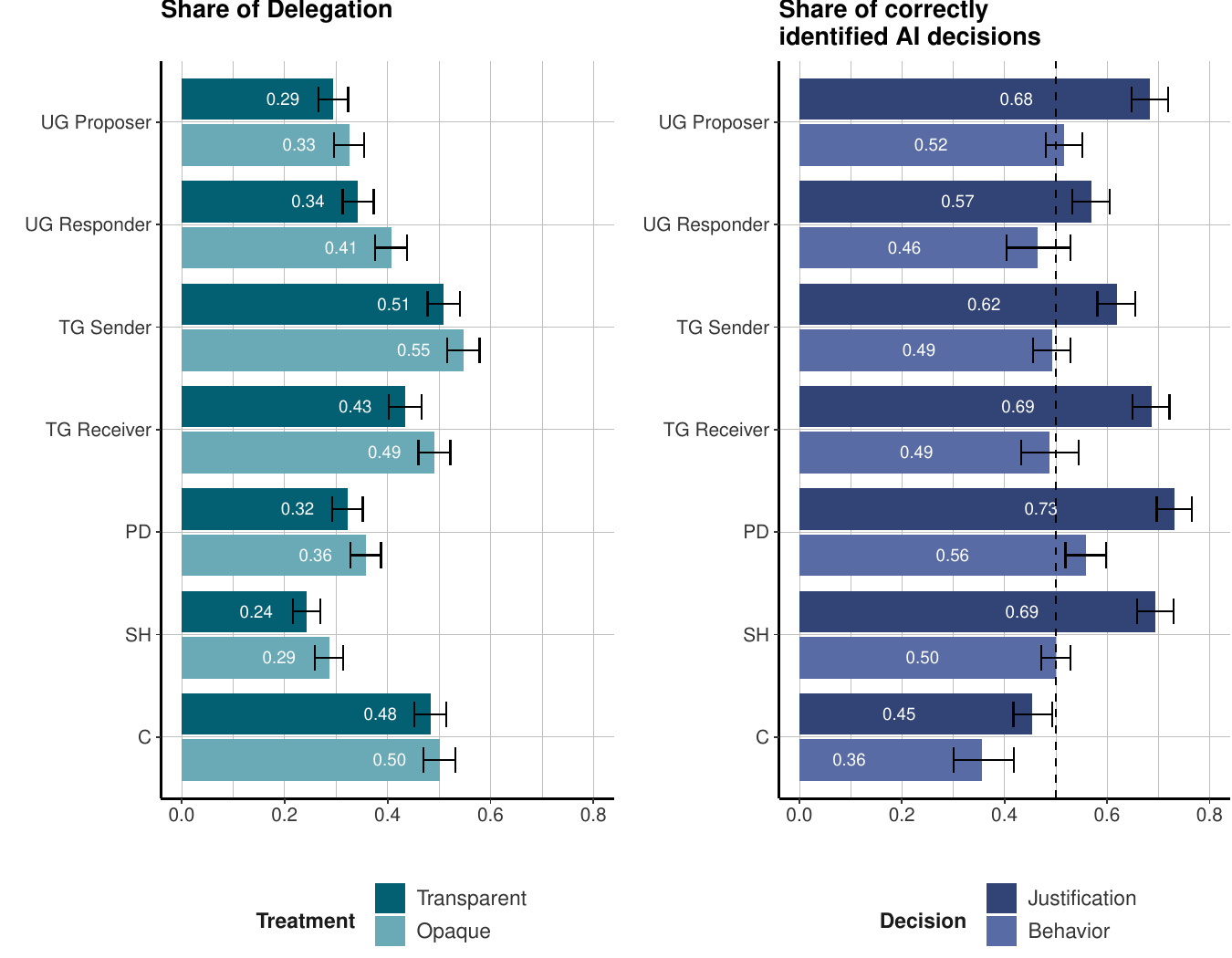}

\caption{Delegation and detectability. Left: share of delegated decisions in treatments \textit{transparent delegation} (light green bars) and \textit{opaque delegation} (dark green bars). Right: Share of correctly identified AI decisions of human raters when seeing pairs of AI and human decisions with (dark blue bars) and without statements of justification (light blue bars). Whiskers show bootstrapped 95\% confidence intervals.}\label{fig:delegation_turing}

\end{figure}

The left panel of Figure~\ref{fig:delegation_turing} shows that at least 20\% of all decisions are delegated to AI in each game-specific situation. This is remarkable because the decisions have consequences. If participants delegate, they do not have to write a justification for the decisions, which creates an incentive to delegate.
Additionally, delegation can be used strategically to avoid being blamed for selfish decisions. Decisions in the stag-hunt game are least often delegated (transparent: 24\%, opaque: 29\%) while the binary trust decision is delegated over 50\% of the time (transparent: 51\%, opaque: 55\%). This may be explained by the fact that cooperation seems to be very attractive in the stag-hunt game (M = 0.81, SD = 0.39) and participants do not want to risk that the AI does not cooperate. Contrasting the frequencies shown in the left panel of Figure~\ref{fig:delegation_turing} to participants' answers of the post-experimental questionnaire, we find that participants anticipate the actual delegation behavior well by stating that others' will delegate to AI (belief in delegation (0-4): M = 2.86, SD = 0.84, comparison to indifference (2), t(1947) = 45.51, p $<$ 0.001).

Figure~\ref{fig:delegation_turing} also shows that participants consistently delegate more if delegation is opaque (transparent delegation: M = 0.38, SD = 0.29, opaque delegation: M= 0.42, SD = 0.29, t(1944) = 3.14, p = 0.006). In all games except the coordination game, opaque delegation is more frequent than transparent delegation (p-values in all situations except C $<$  0.070), which could indicate that participants anticipate the negative welfare consequences of transparent delegation. 

\subsection{Detectability \& Turing Tests}
We now turn to the question of whether people are able to recognize AI decisions or AI-generated justifications for decisions. In each of the seven decisions, humans chose all available options with positive probability, making it fundamentally difficult to distinguish between human and AI decisions. The right panel of Figure~\ref{fig:delegation_turing} shows that the 647 independent human raters who did not take part in the main experiment are generally unable to detect AI decisions. In all games except the prisoner's dilemma, the success rate of human raters is not statistically better than random guessing (p $>$ 0.05 for all situations except PD) when presented with two decisions, one of which was generated by the AI (light blue bars). In the prisoner's dilemma, the AI always cooperates, which is more risky, while humans sometimes choose to defect. This makes identification of the AI possible (success rate PD: M = 0.56, SD = 0.30, t(212) = 2.86, p = 0.002). In contrast, the success rate of human raters is lower than 50\% in the coordination game (two-sided t-test: p $<$ 0.001), which shows that humans can be misled by seemingly human-like decisions made by AI. 

The dark blue bars in the right panel of Figure~\ref{fig:delegation_turing} reveal that the AI does generally not pass the Turing test if raters are additionally shown the written statements of justification for its decision. In all games except the coordination game, the success rate of human raters is significantly higher than chance (p-values for all situations except C $<$ 0.001). The AI generated justifications are generally more technical, often referring to the expected payoffs of the alternatives, than human justifications, which justify decisions with simple language. This might stem from the fact that the AI's training data related to the games mainly stems from texts that target a scientific audience and consequentially seems to use the same level of technicality in its justifications. The only exception is the coordination game for which human raters cannot detect AI generated content even when having access to written justifications. For the coordination game, the AI justifications are less technical and sometimes even a little arbitrary.

\section{Conclusion}\label{sec13}
Recent advancements in generative AI have paved the way for the convenient delegation of a growing number of decisions to algorithms during everyday social interactions.\footnote{By September 2023, ChatGPT had been used 1.5 billion times, and the use of AI on social media platforms is constantly growing~\citep{similarweb2023}.} The findings of our large-scale experiment suggest that this development may have detrimental effects on key underpinnings of human welfare, including fairness, trust, trustworthiness, cooperation, and coordination. 

The utilization of generative AI in social interactions triggers welfare reducing decisions in humans. This pattern persists across all five canonical games we investigate, resulting in negative consequences for participants' payoffs.\footnote{The implications of our findings are strengthened by the observation that prior experience with ChatGPT does not mitigate the negative welfare effects. Nearly half of our participants indicate having used ChatGPT at least once before the study. This high proportion of users among our participants, in contrast to the general population, rules out the possibility that our results are influenced by a lack of familiarity with ChatGPT in our sample compared to the broader population. Official reports indicate that 8\% of US citizens are ChatGPT users, while this increases to 10\% for the population of the UK~\citep{nerdynav2023}.}
Despite this negative effects, participants consistently opt to delegate their decisions to AI when given the opportunity. This is particularly noteworthy given the widespread availability and often cost-free nature of sophisticated generative AI models. Notably, participants exhibit a higher tendency to delegate their decisions when the use of AI is opaque. This suggests an expectation of potential negative welfare consequences linked to transparent delegation. 

Additionally, we discover that opaque delegation to AI can escape detection under specific conditions. In four out of the five games we examine, decisions made by modern generative AI cannot be distinguished when only decisions are reported. In one game, the accuracy of human raters, tasked with identifying decisions secretly delegated to AI, is even lower than chance. These findings underscore the high quality that modern generative AI has achieved in standardized social interactions, such as those explored in our study. However, when raters are provided with justifications for each decision, written by either a human participant or the AI, they can identify AI decisions more frequently than by chance in four out of five games. This highlights that, despite the advancements, some limitations persist in clandestinely employing generative AI in human interaction.  

Taken together, our results indicate that recently proposed regulations to safeguard against the detrimental effects of generative AI on society may be inadequate if they do not consider human reactions to such technology. The study suggests that the potential risks arising from AI decisions constitute only part of the problem. Undesirable changes in human behavior could exacerbate these risks. 

While the EU AI Act\footnote{https://www.europarl.europa.eu/news/en/headlines/society/20230601STO93804/eu-ai-act-first-regulation-on-artificial-intelligence; last retrieved: 08.12.2023}, one of the first comprehensive AI laws globally, requires transparency regarding content generated by generative AI, our findings suggest that such transparency might have unintended consequences. Despite participants in our experiment believing that others may covertly delegate decisions to AI, we observed no increase in the number of welfare-reducing decisions when the use of generative AI is opaque. This stands in contrast to our findings under transparency. These results suggest a leniency pattern when participants are unaware of whether the other person has delegated decisions to AI. Consequently, the proposition to enhance transparency in AI usage, often advocated to minimize negative effects of generative AI on society, paradoxically results in the most pronounced negative effects on welfare in our study~\citep[also see][]{Ishowo2019,Leib2023}.


\backmatter

\bmhead{Supplementary information}
The accompanying supplementary material can be found here: [TBD]

\bmhead{Acknowledgments}
This research was funded by the Deutsche Forschungsgemeinschaft (DFG, German Research Foundation) under Germany’s Excellence Strategy - EXC 2117 - 422037984.
Ethical aspects of the research project were evaluated and approved by the GfeW Institutional Review Board, Certificate No. Ja89kRho, https://gfew.de/ethik/Ja89kRho. We would like to thank the research group at the Thurgau Institute of Economics and the members of Eawag's Department of Environmental Social Sciences, Martijn Kuller and Ueli Reber in particular, for helpful comments and suggestions.

\newpage 
\section{Methods}\label{sec11}

\paragraph{Study type}
The study combines online data collection of AI decisions with an online experiment involving human participants. AI decisions and justification statements are elicited using the text interface of ChatGPT \citep{ChatGPT} (Version Jan 30, free research preview) prior to the online experiment. Participants of the online experiment are recruited via Prolific (www.prolific.com) and randomly assigned to one of the six between-subject treatments. Raters are recruited online via Prolific.

\paragraph{AI decisions}
We sampled AI decisions together with a statement of justification for all decisions of the experiment using the text interface of ChatGPT, Jan 30 Version (courtesy of OpenAI). The algorithm received the same instructions the participants receive during the experiment and therefore knows the consequences of the actions for both players (in experimental currency units (ECU)). 

We sampled personalized AI decisions for each possible  combination of 7 binary preferences ($2^7 = 128$). The algorithm is instructed to act like a person who has preferences defined by a specific pattern of answers to the 7 binary questions. To communicate the preferences of this person to the algorithm, we use the following wording: “Pretend you are a person who prefers intuition over thoughtfulness, introversion over extraversion, fairness over efficiency, chaos over boredom, selfishness over altruism, novelty over reliability, and truth over harmony. Only answer the question at the end. Do NOT explain your answer.” For non-personalized decisions, we use the sentence: “Pretend you are a person. Only answer the question at the end. Do NOT explain your answer.” After the introductory sentence, the algorithm receives information about the game and the decision. 

After the algorithm made the decision, we asked for a statement of justification with the sentence: “Please provide short, informal justification for the decision.” Before moving to the next decision, we deleted all prior conversations and restarted the text interface. In rare cases, in which the algorithm did not provide a valid answer or justification, we deleted the prior conversation, restarted the text interface, and repeated the procedure outlined above.

\paragraph{Experimental games}
Participants of the online experiment play five incentivized economic games. The payoffs are in Experimental Currency Units (ECU) and are  converted to British pounds for payment. \medskip\\

\noindent Ultimatum game (UG) \\
UG proposer: Endowment of 10 ECU. Offer (0-10 ECU) \\
UG responder: Minimum acceptable offer (0-10 ECU) \smallskip\\

\noindent Binary trust game (bTG) \\
bTG sender: Endowment of 5 ECU. Binary decision to send 2 ECU (tripled). \\
bTG trustee: Endowment of 5 ECU. Return in case of trust (0-6 ECU)  \smallskip\\

\noindent Prisoner’s dilemma game (PD) \\
Both C, both get 5 ECU. Both D, both get 3 ECU. \\
You C and the other D, you get 1 ECU and the other gets 8 ECU. \\
You D and the other C, you get 8 ECU and the other gets 1 ECU. \smallskip\\

\noindent Stag-hunt game (SH) \\
Both C, both get 8 ECU. Both D, both get 4 ECU. \\
You C and the other D, you get 1 ECU and the other gets 5 ECU. \\
You D and the other C, you get 5 ECU and the other gets 1 ECU. \smallskip\\

\noindent Coordination game (C)  \\
The five options are: \\
mercury, venus, earth, mars, saturn \\
Same option, 5 ECU. Different options, 2 ECU. \smallskip\\

For the second-mover decisions in the ultimatum game and in the binary trust game we use the strategy method and ask participants for the minimum acceptable offer and the conditional back-transfer in the case of trust. In total, there are 7 different situations in the online experiment (UG sender, UG responder, bTG sender, bTG trustee, PD player, SH player, C player). 

Participants encounter each situation twice, first as the person with AI support and then as the person without AI support. In case of AI support, depending on the treatment, the participants either have to select a decision for the case that AI does not take over or they have to decide whether to delegate the decision to AI. Without AI support, depending on the treatment, the participants either have to select a decision for each potential and known interaction partner - AI or human - or select one decision for an unknown interaction partner.

\paragraph{Experimental treatments}
The experiment uses a between-subject design with six treatments. In the first treatment condition, AI makes the decision of the participant with AI support with 50\% probability (\textit{transparent random}). Participants without AI support make two decisions, one for the case of interacting with a human, and one for the case of interacting with AI – which makes the use of AI transparent. In the second treatment condition, the participant with AI support can choose if she wants to delegate her decision to AI (\textit{transparent delegation}). In this condition, the use of AI is also transparent as the participant without AI support makes two decisions, one for each outcome of the delegation decision. The third treatment condition is a variant of the delegation treatment in which participants without AI support cannot condition their decision on the outcome of the delegation decision, which makes the use of AI opaque (\textit{opaque delegation}). For each of these three treatment conditions, we vary whether participants can personalize the decisions AI makes on their behalf at the beginning of the experiment (\textit{personalized}) or not (\textit{non-personalized}). 

This results in six between-subject treatments overall: 
\begin{enumerate}
    \item TRN: transparent random non-personalized
    \item TRP: transparent random personalized
    \item TDN: transparent delegation non-personalized
    \item TDP: transparent delegation personalized
    \item ODN: opaque delegation non-personalized
    \item ODP: opaque delegation personalized
\end{enumerate}

Each participant is randomly assigned to one of the six treatment conditions: TRN, TRP, TDN, TDP, ODN, and ODP. In the treatments with random takeover of AI (TRN, TRP), AI takes over the decision of the AI supported participant with 50\% probability.

In the transparent treatments (TRN, TRP, TDN, TDP), participants decide separately for the scenario of interacting with a human or the scenario of interacting with AI. In the opaque delegation treatments (ODN, ODP), participants are not aware of the delegation decision and therefore do not know if they interact with a human or with AI. 

\paragraph{Monetary incentives}
After a participant has completed all 14 decisions, she is randomly matched to another participant from the same treatment. For each pair of participants, the experimenter randomly selects one of the 14 interactions, which determines the game and the players’ roles in the game, and randomly determines which participant is supported by AI. If the treatment involves the possibility of random takeover by AI (TRN, TRP), the experimenter randomly determines if AI makes the decision on behalf of the participant with AI support. The payoffs of both participants are then determined using the decisions of both players (human or AI) in the selected interaction. There is no feedback between interactions.

\paragraph{Personalization of AI}
In the treatments with personalization, we use participants’ answers to 7 binary preference questions to personalize AI decisions made on their behalf, which results in 128 possible answer patterns. Each unique pattern of answers reflects a certain personality of the AI. The questions follow the simple format “A or B?” and are: intuition or thoughtfulness, introversion or extraversion, fairness or efficiency, chaos or boredom, selfishness or altruism, novelty or reliability and, truth or harmony.

Participants know the purpose of the questions - that their answers will influence the decisions AI makes on their behalf in the online experiment. Participants do not know the instructions of the games at the time when answering the binary questions, which limits (but does not exclude) strategic personalization. At the time when participants personalize the AI, they only know that they will interact with other participants and that their decisions will have consequences for both participants. 

\paragraph{Experimental procedures}
All participants receive general instructions with information about the study and the experimental procedures and are subsequently asked to give informed consent. Participants then receive detailed information about the AI model used in the online experiment. We provide examples of model output in tasks unrelated to the economic games studied in the experiment. We then ask for participants’ prior experience with ChatGPT. In the treatments with personalized AI, we provide an additional example for the effect of personalizing the response of ChatGPT. Participants are then asked 7 binary questions in the simple format “A or B?” to personalize decisions made by AI on their behalf. This happens before participants receive the instruction for the games to limit the possibility of strategic personalization.  

Participants receive general instructions on how the interaction with the AI model works (treatment specific but not game specific) and answer several control questions including two attention checks. Participants then receive the instructions for the first game. We use standardized instructions for all games~\citep{Thielmann2021,Mehta1994} with minor modifications for SH and C. Participants are informed about their role in the game and are asked for their decision(s) with AI support.

For all decisions except those that are delegated to AI, participants write a short statement of justification immediately after making the decision. Participants know that the justifications will not be shown to other participants during the experiment. Participants are subsequently asked for their decision(s) as participant without AI support and their statement(s) of justification. If the game has more than one player role, the procedure is repeated for the other role. No feedback about the decisions of other participants or the AI decisions is provided between the interactions. 

Participants with AI support make one decision in the treatments with random AI takeover (TRN, TRP) for the case that AI does not make the decision. In all treatments with delegation (TDN, TDP, ODN, ODP) participants with AI support first decide whether they would like to delegate and are subsequently only asked for their decision if they do not delegate to AI. Participants in the transparent treatments (TRN, TRP, TDN, TDP) make two decisions in the interactions without AI support, one for the case of interacting with a human and, one for the case of interacting with AI.  Participants in the opaque delegation treatments (ODN, ODP) make one unconditional decision in interactions without AI support.

After a participant has completed all games in a randomized order, we ask participants for their age, gender, and education level. We also ask each participant to rate the predictability and kindness of AI, the importance of equality and intentions when interacting with AI, the resemblance of AI decision to human preferences, and in the delegation treatments the belief in and the appropriateness of delegation to AI on a 5-point Likert scale. After the experiment, each participant is matched to another participant from the same treatment by the experimenter. The experimenter randomly selects one of the 14 interactions for each pair and the matched participants receive the payoffs resulting from this interaction in addition to a flat payment for participation. 

\paragraph{Detectability \& Turing tests}
For each treatment condition, we create several collections of decisions, each consisting of 7 human decisions and 7 AI decisions. To generate the collections, we filter out all decisions with statements revealing that (1) a human made the decision or (2) AI made the decision. This includes, for example, all human decisions with statements that reference information not available to AI (e.g., information about the general procedure of the experiment, the fact that the participant is interacting with AI) and AI statements revealing AI generated decisions. We also filter out human statements with less than 3 words and correct the selected human statements for typos or grammatical errors. 

For the data of four treatments TRN, TRP, TDN, and TDP, each rater compares the 7 decisions a participant made without AI support targeted at a human to the 7 AI decisions generated for the same participant in the same interaction and indicates for each situation which decision was generated by AI. All decisions are presented together with the corresponding statements of justification written by the participant or written by AI. The rater receives a bonus payment if her rating in one randomly selected situation is correct. 

For the data of the treatments ODN and ODP, the raters compare decisions of participants with AI support who do not delegate their decision to the decision generated by AI for these participants in the same situation. We collect 14 ratings from each rater. Each rater reviews 7 decisions, one from each of the 7 situations in the online experiment. First, we present a pair of decisions without the written statement of justification and ask the rater to decide which decision AI made. Then, we reveal the written statements of justification and give the rater the opportunity to revise her rating. At the end, one situation and one of the two ratings is randomly selected and the rater receives a bonus payment if the rating is correct.

For the Turing tests (exploratory hypotheses H9 and H10), we are interested in the share of correctly identified AI decisions. In the treatments ODN and ODP we post-hoc randomize the positioning of the AI decisions in situations in which the AI and the human actions were identical to exclude any ordering bias. We do so by setting the outcome variable of correctly identifying the AI decision to 0.5. We also analyze participants’ Likert-scale ratings about the nature of human-AI interaction and socio-demographic variables collected after the experiment. We also measure response times and how often a participant leaves or switches tabs in the browser (or cannot see the browser window) during the online experiment. 

\paragraph{Ethical approval and informed consent} 
The online experiment was carried out in accordance with the  regulations of the ethics committee of the University of Konstanz. All experimental protocols were approved by the institutional review board of Gesellschaft f\"{u}r Experimentelle Wirtschaftsforschung (GfEW). Institutional Review Board Certificate No. Ja89kRho (https://gfew.de/ethik/Ja89kRho). Informed consent was obtained from all participants prior to participation.

\paragraph{Sample size}
The target sample size was 3000 experimental subjects in the main experiment (500 per treatment) and 600 human raters (100 per treatment) for the Turing tests. We reached a sample size of 2905 experimental subjects (mean age: 39.9 years, 49.6\% women, 49.5\% men, 0.7\% non-binary, 47.1\% college or university degree) and 647 human raters (mean age: 39.8 years, 50.1\% women, 48.8\% men, 1.1\% non-binary, 43\% college or university degree).
We estimated the required sample size of each treatment based on a simulation. For each of our main hypotheses (H1-H8), we simulated data with effects of 5 ppt in the direction of the alternative hypothesis for each decision affected by the alternative hypothesis. Rejecting H1-H8 with probability $\ge$ 0.90 at an alpha level of 5\%, using Holm-Bonferroni corrected p-values for multiple testing, required a sample size of approximately 500 participants per treatment. 

As a technical check, we ran a pilot experiment online with 60 participants recruited via Prolific. 

The median duration time in the first experiment was 17 minutes and participants earned 7 pounds per hour on average. In the second experiment, the median duration time was 8.5 minutes with average earnings of 14 pounds per hour.

\paragraph{Experimental variables}
We observe participants' decisions in 14 different interactions with another unknown participant from the same treatment. Our key variables of interest are the decisions participants make in the 5 two-player games, which are: the offer in UG, the minimum acceptance threshold in UG, the binary trust decision in bTG, the conditional back-transfer in case of trust in bTG, the binary cooperation decision in PD, the binary cooperation decision in SH, and the selection of the modal choice in C (earth).  

We generate the following normalized variables:
\begin{itemize}
    \item Normalized offer in UG = offer/5  \\ 
Offers are usually between 1 and 5, offers larger than 5 are truncated.
\item Normalized min acceptance threshold in UG = min/5  \\
Thresholds are usually between 1 and 5, thresholds larger than 5 are truncated.
\item Normalized back-transfer in bTG = back-transfer/6  \\
Back-transfers are between 0 and 6 (no truncation needed).
\end{itemize}

\paragraph{Data exclusion}
We discard the observations of participants who report that they have been repeatedly disturbed and therefore recommend not to use their data in the analysis in an open question asked at the end of the experiment. Participants are informed that not using their data has no payoff consequences for them. We also discard the observation of participants who state in an open question at the end that they consulted ChatGPT for advice during the experiment. We also discard all decisions with unreasonable response times (either too fast or too slow). A decision is discarded if the logarithm of the time needed to make the decision is more than 3 standard deviations away from the mean.

\paragraph{Research Questions and Hypotheses}
We derive 8 hypotheses (H1--H8) related to 8 research questions (Q1-Q8) derived from five general assumptions about the nature of modern AI, which are outlined in the pre-analysis plan.\medskip\\

\noindent Q1: \textit{How does the use of AI change social interaction?}\\
We compare the decisions targeted at humans to the decisions of the same participants targeted at AI in TRN.\\
H1: In TRN, the welfare index is smaller for decisions targeted at AI.
\smallskip\\

\noindent Q2: \textit{Do people delegate more if delegation is opaque?}\\
We compare the frequency of delegation in the pooled data of TDN and TDP to the frequency of delegation in pooled data of ODN and ODP.\\
H2: Delegation is more frequent in ODN and ODP compared to the pooled data of TDN and TDP.\smallskip\\

\noindent Q3: \textit{Does the possibility of opaque delegation change social interaction?}\\
We compare the decisions targeted at humans in TRN to the unconditional decisions of participants in ODN (who do not know if they interact with a human or with AI).\\
H3: The welfare index is smaller for decisions in ODN than in TRN.\smallskip\\

\noindent Q4: \textit{Does delegation to AI crowd-out prosocial behavior?}\\
We compare the decisions targeted at AI between TDN and TRN.\\
H4: Prosociality index is smaller for decisions in TDN than in TRN.\smallskip\\

\noindent Q5: \textit{Does social interaction change less if AI is personalized?}\\
We compare the decisions targeted at AI between TRP and TRN.\\
H5: The welfare index is larger for decisions in TRP than in TRN.\smallskip\\

\noindent Q6: \textit{Do people delegate more to personalized AI?}\\
We compare the frequency of delegation in the pooled data of TDN and ODN to the frequency of delegation in the pooled data of TDP and ODP.\\
H6: Delegation is more frequent in pooled data of TDP and ODP compared to the pooled data of TDN and ODN.\smallskip\\

\noindent Q7: \textit{Does the possibility of opaque delegation to personalized AI change social interaction less?}\\
We compare the unconditional decisions of participants in ODP and ODN.\\
H7: The welfare index is larger for decisions in ODP than in ODN.\smallskip\\

\noindent Q8: \textit{Does delegation to personalized AI reduce the crowding-out of prosocial behavior?}\\
We compare the decisions targeted at AI between TDP and TDN.\\
H8: The prosociality index is larger for decisions in TDP than in TDN.\bigskip\\

\paragraph{Statistical analyses}
We use a conditional testing procedure to correct for multiple testing. First, we test H1--H8 using Holm-Bonferroni corrected p-values, which controls for the fact that we initially perform 8 tests using an alpha level of 5\%. If a test is significant, because its corrected p-value is smaller than 5\%, we perform conditional tests for the same data.  For hypotheses about welfare (H1, H3, H5, H7), we additionally test for differences in the predictability of behavior (frequency of modal choice in C), equality concerns (offers in UG), the kindness index, and the intentions index (4 conditional tests). 

For hypotheses about prosociality (H4, H8), we test for differences in the 6 normalized decisions underlying the prosociality index: the normalized offer in the ultimatum game (offer/5), the normalized minimum acceptance threshold (min/5), the binary trust decision, the normalized back-transfer in the trust game (back-transfer/6), and the binary cooperation decision in the prisoner’s dilemma and the binary cooperation decision in the stag-hunt game (6 conditional tests). For hypotheses about delegation (H2, H6), we test if the frequency of delegation differs in each game-specific situation (7 conditional tests). Simulation results suggest that the conditional testing procedure effectively restricts the familywise error rate. The simulated probability of at least one Type I error is 4.5\% for the main tests and 4.3\% for the conditional tests (estimated based on 10,000 simulation runs).

\paragraph{Preregistration}
The study was preregistered. The preregistration files of the study can be found here: https://osf.io/fvk2c/

\newpage 

\appendix

\begin{appendices}

\renewcommand\theHtable{Appendix.\thetable}

\section{}\label{secA1}

\begin{table}[!htbp] \centering 
  \caption{Influence of covariates on the (loss of) welfare} 
  \label{tab:regression} 
\begin{tabular}{@{\extracolsep{5pt}}lccc|ccc} 
\toprule
 & \multicolumn{6}{c}{\textit{Dependent variable:}} \\ 
& \multicolumn{3}{c|}{Welfare Index (base: AI)} & \multicolumn{3}{c}{Diff. in Welfare Index (Human-AI)} \\ 
& (1) & (2) & (3) & (4) & (5) & (6)\\ 
\midrule
\textit{Decision targeted at}... & \\
...Human & 0.060$^{***}$ & 0.060$^{***}$ & 0.060$^{***}$ &  &  &  \\ 
  & (0.004) & (0.004) & (0.004) &  &  &  \\ 
...Unknown & 0.054$^{***}$ & 0.049$^{***}$ & 0.046$^{***}$ &  &  &  \\ 
  & (0.007) & (0.007) & (0.007) &  &  &  \\ 
 Age &  & 0.0004 & 0.001$^{**}$ & $-$0.001$^{*}$ & $-$0.0003 & $-$0.001$^{*}$ \\ 
  &  & (0.0002) & (0.0002) & (0.0003) & (0.0003) & (0.0003) \\ 
 Gender: Non-binary &  & $-$0.037 & $-$0.035 & 0.101$^{**}$ & 0.085$^{**}$ & 0.081$^{*}$ \\ 
  &  & (0.034) & (0.034) & (0.043) & (0.042) & (0.042) \\ 
 Gender: Woman &  & $-$0.014$^{**}$ & $-$0.014$^{**}$ & $-$0.001 & $-$0.003 & $-$0.003 \\ 
  &  & (0.006) & (0.006) & (0.008) & (0.008) & (0.008) \\ 
 Higher Education &  & $-$0.008 & $-$0.008 & 0.001 & 0.001 & 0.0005 \\ 
  &  & (0.007) & (0.007) & (0.009) & (0.008) & (0.008) \\ 
 Usage of ChatGPT&  & 0.004 & 0.004 & 0.0001 & 0.002 & 0.002 \\ 
  &  & (0.003) & (0.003) & (0.004) & (0.004) & (0.004) \\ 
\textit{Agreement to}... & \\
...AI difficult to predict &  & $-$0.006$^{*}$ & $-$0.006$^{*}$ &  & 0.017$^{***}$ & 0.017$^{***}$ \\ 
  &  & (0.003) & (0.003) &  & (0.004) & (0.004) \\ 
...AI trustworthy &  & 0.015$^{***}$ & 0.016$^{***}$ &  & $-$0.036$^{***}$ & $-$0.036$^{***}$ \\ 
  &  & (0.004) & (0.004) &  & (0.005) & (0.005) \\ 
...Equality concern with AI &  & 0.055$^{***}$ & 0.054$^{***}$ &  & $-$0.015$^{***}$ & $-$0.014$^{***}$ \\ 
  &  & (0.003) & (0.003) &  & (0.004) & (0.004) \\ 
...AI reflects human &  & $-$0.002 & $-$0.002 &  & $-$0.001 & $-$0.001 \\ 
  &  & (0.003) & (0.003) &  & (0.003) & (0.003) \\ 
 Distractions &  &  & 0.00001 &  &  & $-$0.001 \\ 
  &  &  & (0.001) &  &  & (0.001) \\ 
 Duration (min) &  &  & $-$0.001$^{***}$ &  &  & 0.002$^{***}$ \\ 
  &  &  & (0.0004) &  &  & (0.0005) \\ 
 Constant & 0.649$^{***}$ & 0.472$^{***}$ & 0.486$^{***}$ & 0.080$^{***}$ & 0.136$^{***}$ & 0.119$^{***}$ \\ 
  & (0.004) & (0.019) & (0.020) & (0.016) & (0.024) & (0.025) \\ 
\midrule
Observations & 4,808 & 4,808 & 4,808 & 1,903 & 1,903 & 1,903 \\ 
R$^{2}$ & 0.024 & 0.138 & 0.140 & 0.005 & 0.062 & 0.067 \\ 
Adjusted R$^{2}$ & 0.023 & 0.136 & 0.137 & 0.002 & 0.057 & 0.062 \\ 
\bottomrule
\end{tabular} 
\footnotetext{Note: OLS regressions with welfare index (baseline: decisions targeted at AI) observed in all treatments as dependant variable in models (1)--(3) and difference in welfare index (human - AI) observed in the treatments \textit{transparent random} as dependant variable in models (4)--(6).\textit{Usage of ChatGPT} is the numeric representation of the frequency of the use of ChatGPT. The four variables of agreement to specific statements about the AI are measured on a scale from 0--4. \textit{Distractions} is the frequency of changing the browser window during the experiment. Robust standard errors clustered at the subject level in models (1)--(3). $^{*}$p$<$0.1; $^{**}$p$<$0.05; $^{***}$p$<$0.01}
\label{tab:regression}
\end{table} 

\end{appendices}
\newpage

\bibliography{sn-bibliography}

\end{document}